\documentclass[a4paper,11pt]{article}

\usepackage{hyperref}
\usepackage{amsmath,amsthm,amssymb}
\usepackage{xcolor}
\usepackage{graphicx}
\usepackage{epstopdf}
\usepackage{overpic}
\usepackage{bbold}
\usepackage{textcomp}
\usepackage{siunitx}
\usepackage{physics}
\usepackage{subcaption}
\usepackage{mathtools}

\usepackage{tikz}
\usepackage{circuitikz}
\usetikzlibrary{shapes.misc}
\usetikzlibrary{colorbrewer}
\tikzset{cross/.style={cross out, draw=black, fill=none, minimum size=2*(#1-\pgflinewidth), inner sep=0pt, outer sep=0pt}, cross/.default={2pt}}

\usepackage{pgfplots}
\pgfplotsset{compat=1.10}
\usepgfplotslibrary{colorbrewer}
\pgfplotsset{cycle list/Set1-5}

\usepackage{authblk}

\renewcommand{\vec}[1]{\boldsymbol{#1}}

\newcommand{\dx}{\;\text{d}x}
\newcommand{\dt}{\;\text{d}t}
\newcommand{\dA}{\;\text{d}\vec{A}}

\newcommand{\dV}{\;\text{dV}}

\begin{document}

\title{Micromagnetic Simulation and Optimization of Spin-Wave Transducers}

\author[1]{Florian Bruckner \thanks{florian.bruckner@univie.ac.at}}
\author[1]{Krist\'yna Dav\'idkov\'a}
\author[1]{Claas Abert}
\author[1]{Andrii Chumak}
\author[1]{Dieter Suess}

\affil[1]{Faculty of Physics, University of Vienna, Austria}
\maketitle

\begin{abstract}
The increasing demand for higher data volume and faster transmission in modern wireless telecommunication systems has elevated requirements for 5G high-band RF hardware.
Spin-Wave technology offers a promising solution, but its adoption is hindered by significant insertion loss stemming from the low efficiency of magnonic transducers.
This work introduces a micromagnetic simulation method for directly computing the spin-wave resistance, the real part of spin-wave impedance, which is crucial for optimizing magnonic transducers.
By integrating into finite-difference micromagnetic simulations, this approach extends analytical models to arbitrary transducer geometries.
We demonstrate its effectiveness through parameter studies on transducer design and waveguide properties, identifying key strategies to enhance the overall transducer efficiency.
Our studies show that by varying single parameters of the transducer geometry or the YIG thickness, the spin-wave efficiency, the parameter describing the efficiency of the transfer of electromagnetic energy to the spin wave, can reach values up to 0.75. The developed numerical model allows further fine-tuning of the transducers to achieve even higher efficiencies.
\end{abstract}

\section{Introduction}
Magnonics explores the excitation, propagation, control, and detection of spin waves \cite{barman20212021, chumak2022advances}. Unlike conventional logic circuits, magnonic systems reduce energy loss due to the absence of ohmic dissipation and offer superior CMOS compatibility over photonics and phononics \cite{mahmoud2020introduction}. The field of magnonics is advancing rapidly, with key developments in 3D architectures \cite{fernandez2017three}, magnonic logic gates for Boolean processing \cite{kostylev2005spin}, neuromorphic computing \cite{papp2021nanoscale}, magnetic sensors \cite{gattringer2023offset}, and quantum magnonics leveraging hybrid entangled states \cite{lachance2019hybrid}. A particularly promising direction is magnonic RF applications \cite{dieny2020opportunities,levchenko2024review,davidkova2025nanoscaled}, driven by growing 5G technology demands \cite{agiwal2016next}. Progress in these fields relies heavily on efficient transducers between electrical signals or electromagnetic waves with spin-wave signals.

Radio frequency (RF) spin-wave devices, based on micrometer-thick films and lateral dimensions on the centimeter scale, achieve insertion losses below \SI{3}{dB} \cite {levchenko2024review}. However, the miniaturization of modern devices necessitates reducing the thickness of YIG films to the nanoscale \cite{dubs2020low} and lateral dimensions to the range of hundreds of micrometers. Current transducers for nanoscale spin-wave RF filters and frequency-selective limiters, however, exhibit insertion losses exceeding \SI{20}{dB} \cite{davidkova2025nanoscaled}.

The key quantity for transducer optimization is the radiation impedance $Z_r$ and especially the radiation resistance $R_r$ which is associated with the power of the excited spin wave.
We propose to use the term "spin-wave impedance" instead of "radiation impedance" to clearly distinguish it from the electromagnetic radiation impedance commonly referred to in classical antenna theory.
Thus, in this paper we use the terms "spin-wave resistance" $R_\text{sw}$ and "spin-wave efficiency" $\eta_\text{sw}$. Analytical models have been developed for the spin-wave resistance in various configurations, including Damon-Eshbach\cite{ganguly1975microstrip}, Backward-Volume\cite{parekh1980excitation}, and Forward-Volume\cite{parekh1979theory}. Effects such as ohmic losses and self-inductance can be incorporated using a lumped-circuit model, enabling the simulation of magnetic devices as equivalent electrical circuits \cite{vanderveken2022lumped}.

Instead of relying on analytical calculations as in \cite{connelly2021efficient}, the proposed approach determines the spin-wave resistance using micromagnetic simulations, which allow arbitrary shapes and configurations and provide a generalized framework for the optimization of magnonic transducers. Micromagnetic simulations have been successfully applied in a wide range of applications, including magnetic storage \cite{fidler2000micromagnetic}, spintronics \cite{abert2019micromagnetics}, sensing \cite{suess2018sensor}, and the optimization of magnetic circuits and materials \cite{wang2024nanoscale}. In magnonics, they have led to breakthroughs such as inverse-design units \cite{wang2021inverse,kiechle2023spin,zenbaa2024magnonic}, and integrated magnonic circuits \cite{wang2020coupler}.

A similar approach for the combination of circuit-level models with micromagnetic simulations \cite{erdelyi2024design} is based on a combination of simulation tools, e.g. mumax$^3$ for the micromagnetic part, and time harmonic magnetic FEMM simulations for solving the excitation field created by the transducer. In contrast to this, the presented method perfectly integrates into existing finite-difference micromagnetic simulations. The implementation using a single \textit{magnum.np} \cite{bruckner2023magnum} script simplifies the simulations process and opens the possibility to utilize modern optimization approaches like inverse design.

\section{Method}
\begin{figure}[h!]
    \centering
    \begin{tikzpicture}[american, scale=0.8, transform shape]
        \draw (0,2) to[short, o-, i^=$i_1(t)$] (0.5,2)
                    to[R, l=$R_\Omega$] (2.0,2)
                    to[L, l=$L_0$] (3.5,2)
                    to[R, european, l=$Z_{11}$] (5.0,2)
                    to[short] (5.5,2)
                    to[controlled voltage source, v_=$Z_{12}i_2$] (5.5,0)
                    to[short,-o] (0,0);

%

        \draw[->, thick, >=Stealth] (0,1.7) -- (0,0.3) node[midway,left] {$u_1(t)$};

        \node at (6.25,1.1) {$\vec{\sim}$};
        \node at (6.25,1.0) {$\vec{\sim}$};
        \node at (6.25,0.9) {$\vec{\sim}$};

        \begin{scope}[opacity=0.3]
        \draw (12.5,0) to[short,o-] (7,0)
                       to[controlled voltage source, v_=$Z_{21} i_1$, invert] (7,2)
                       to[short] (7.5,2)
                       to[R, european, l=$Z_{22}$] (9.0,2)
                       to[L, l=$L_0$] (10.5,2)
                       to[R, l=$R_\Omega$] (12.0,2)
                       to[short, -o, i^<=$i_2(t)$] (12.5,2);
        \draw[->, thick, >=Stealth] (12.5,1.8) -- (12.5,0.2) node[midway,right] {$u_2(t)$};
        \end{scope}

        \draw[dashed, thick, Set1-B] (3.5,-0.3) rectangle (9.0,2.8);
        \node[blue, label=below:\textcolor{Set1-B}{Magnonic System}] at (6.25, -0.3) {};
    \end{tikzpicture}
    \caption{Equivalent circuit representation of a magnonic filter consisting of a primary and a secondary transducer, with the ohmic resistance $R_\Omega$ and the inductance $L_0$, separated by a magnonic waveguide. The magnonic system is described as a two-port network with impedance parameters $Z_{11}$, $Z_{12}$, $Z_{21}$, and $Z_{22}$. For the optimization of the overall efficiency of the filter the main quantity is the ratio of the spin-wave resistance $R_\text{sw} = \Re{Z_{11}}$ and the ohmic resistance $R_\Omega$.}
    \label{fig:circuit}
\end{figure}
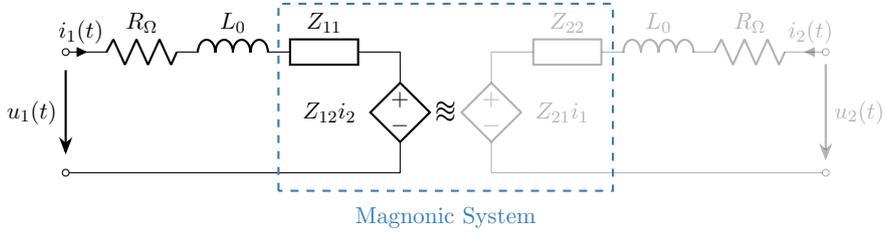

The optimization of spin wave transducers requires quantifying the energy transferred from the electric power source to the magnetic spin wave.
Since typical magnon wavelengths $\lambda^\text{sw}$ are much smaller than the corresponding electromagnetic wavelength $\lambda^\text{em} = c / f$, where $c$ is the speed of light and $f$ is the frequency, the magnonic transducer can be modeled using a lumped circuit representation. For instance, at a frequency of \SI{25}{GHz}, the electromagnetic wavelength would be approximately \SI{1}{cm}. As proposed by \cite{erdelyi2024design}, the equivalent circuit shown in Fig.~\ref{fig:circuit} describes the magnonic system, which consists of a primary transducer that excites the spin wave, a magnetic waveguide that guides the spin wave, and a secondary transducer that measures the induced electrical signal.

Assuming an impressed current $i_1(t) = \hat{I}_1 \sin(2 \pi f t)$ the power which is converted into a spin wave $P_\text{sw}$ and the ohmic losses $P_\Omega$ can be expressed as
\begin{align}
\begin{split}
    P_\text{sw} &= \frac{1}{2} R_\text{sw} \hat{I}_1^2 \\
    P_\Omega &= \frac{1}{2} R_\Omega \hat{I}_1^2,
\end{split}
\end{align}
where $R_\text{sw} = \Re{Z_{11}}$ and $\hat{I}_1$ is the peak value of the current $i_1(t)$. This allows to define the following spin-wave efficiency
\begin{align}
    \eta_\text{sw} = \frac{R_\text{sw}}{R_\text{sw} + R_\Omega} = \frac{P_\text{sw}}{P_\text{sw} + P_\Omega}
\end{align}

Furthermore, parts of the input power will be reflected at the transducer if its impedance $Z_T = R_\Omega + i \omega L_0 + Z_{11}$ is not matching the characteristic impedance of the source or the transmission line connected to the transducer $Z_0$ (typically \SI{50}{\ohm}).
This mismatch gives rise to the following matching efficiency
\begin{align*}
    \eta_\text{m} = 1-\Gamma^2, \quad \text{with} \quad \Gamma = \frac{Z_T-Z_0}{Z_T+Z_0}
\end{align*}

Finally, one ends up with the following transducer efficiency $\eta_T = \eta_\text{sw} \; \eta_\text{m}$ which will be used as a figure of merit for the following transducer optimizations. It represents the proportion of power transferred from the electrical signal to the spin wave.

While the ohmic resistance $R_\Omega$ can be analytically calculated, micromagnetic simulations are used for the calculation of the spin-wave resistance $R_\text{sw}$.
The time-domain Landau-Lifshitz-Gilbert(LLG) equation is solved using \textit{magnum.np} \cite{bruckner2023magnum}, a finite-difference micromagnetic simulation tool based on PyTorch
\begin{align}\label{eqn:llg}
    \dot{\vec{m}} = -\frac{\gamma}{1+\alpha^2} \left[ \vec{m} \times \vec{h}^\text{eff} + \alpha \, \vec{m} \times \left( \vec{m} \times \vec{h}^\text{eff} \right) \right],
\end{align}
with the reduced magnetization $\vec{m}$, the gyromagnetic ratio $\gamma = \SI{2.21e5}{m/As}$, and the dimensionless damping constant $\alpha$. The effective field $\vec{h}^\text{eff} = \vec{h}^\text{ex} + \vec{h}^\text{d} + \vec{h}^\text{u} + \vec{h}^\text{oe} + \vec{h}^\text{bias}$ incorporates contributions from exchange field $\vec{h}^\text{ex}$, uniaxial anisotropy field $\vec{h}^\text{u}$, demagnetization field $\vec{h}^\text{d}$, external bias field $\vec{h}^\text{bias}$, as well as from the Oersted field $\vec{h}^\text{oe}$, resulting from an impressed current density $\vec{j}$.

For a certain transducer geometry and a corresponding spatial current density the high-frequency excitation leads to an Oersted field of the following form
\begin{align}
\begin{split}
    \vec{h}^\text{oe}(\vec{x}, t) &= \vec{h}^\text{oe}_0(\vec{x}) \; \sin(2 \pi f t) \\
    \vec{h}^\text{oe}_0(\vec{x}) &= \frac{1}{4 \pi} \int \vec{j}(\vec{x}') \times \frac{\vec{x}-\vec{x}'}{\vert \vec{x}-\vec{x}'\vert^3} \, \dx'
\end{split}
\end{align}
The Oersted field can thus be calculated only once for a normalized current density. Within the time-integration, the field is then scaled according to the time-dependent excitation.
Compared to other published methods \cite{erdelyi2024design, connelly2021efficient} \textit{magnum.np} enables the direct evaluation of the Oersted field within a single simulation using an efficient FFT-based approach. This simplifies the simulation process and opens the possibility to utilize modern optimization approaches like inverse design. On the downside {\it magnum.np} currently assumes a homogeneous current density across the antenna ignoring skin-effect and proximity-effect. However, this should be a valid assumption as long as the transducer width is smaller than the skin-depth (e.g. the skin-depth in copper at $f=\SI{4}{GHz}$ is $\delta = \sqrt{\frac{\rho}{\pi \mu f}} \approx \SI{1}{\mu m}$).

For a known peak input current $\hat{I}_1$ the spin-wave resistance corresponding to the excited spin wave can be calculated via the average spin wave power $P_\text{sw}$
\begin{align}
    R_\text{sw} = \frac{2 \, P_\text{sw}}{\vert \hat{I}_1 \vert^2}, \quad \text{with} \quad P_\text{sw} = \frac{1}{T} \int_T p_\text{sw}(t) \dt,
\end{align}
with the time-dependent power $p_\text{sw}$ which is averaged over (multiple) periods $T$.

One way to derive the time-dependent power $p_\text{sw}$ is the numerical time derivative of the micromagnetic energy E, which is easily accessible with most micromagnetic tools.
\begin{align}
    p_\text{sw}(t) = \frac{\partial E}{\partial t} \quad \quad E = -\mu_0 \int M_s \vec{m} \cdot \left( \frac{1}{2} \vec{h}^\text{lin}[\vec{m}] + \vec{h}^\text{bias} \right) \dV
\end{align}

As the spin wave front propagates in the waveguide, the region containing the spin wave is increasing with time and the total micromagnetic energy is increasing.
Alternatively, the time-dependent power could also be calculated via the induced voltage
\begin{subequations}
\begin{align}
    p_\text{sw} = u_\text{ind}(t) \, i_1(t),
\end{align}
which is in turn related to the flux linkage $\psi_m$ through the transducer
\begin{align}
    u_\text{ind}(t) = -\frac{\partial \psi_m}{\partial t}, \quad \quad \psi_m = \mu_0 \int M_s \vec{m}(t) \cdot \hat{\vec{h}}^\text{oe} \dV = \int \vec{B}_m \dA,
\end{align}
where $\hat{\vec{h}}^\text{oe} = \vec{h}^\text{oe}_0(\vec{x}) / \hat{I}_1$ is the normalized spatial profile of the Oersted field created by a transducer and $\vec{B}_m = \mu_0 (M_s \, \vec{m} + \vec{h}^\text{d})$ is the magnetic flux density created by the magnetization. Using the volume integral as proposed by \cite{vanderveken2022lumped} instead of the surface integral simplifies usage and prevents complications for thick transducers, where the chosen location of the cross-section has a strong influence on the resulting induced voltage. Note, that $\psi_m$ only contains the flux contributions which result from the changing magnetization, while effects such as self-inductance are covered by $L_0$.
\end{subequations}



\section{Numerical Experiments}
The presented method is used to investigate the optimization of transducers used for recently published frequency selective limiters \cite{davidkova2025nanoscaled}. The current study focuses on a transducer in the Damon-Eshbach configuration at an operating frequency $f=\SI{4}{GHz}$.

\begin{figure}[h!]
  \centering
  \includegraphics[width=10cm]{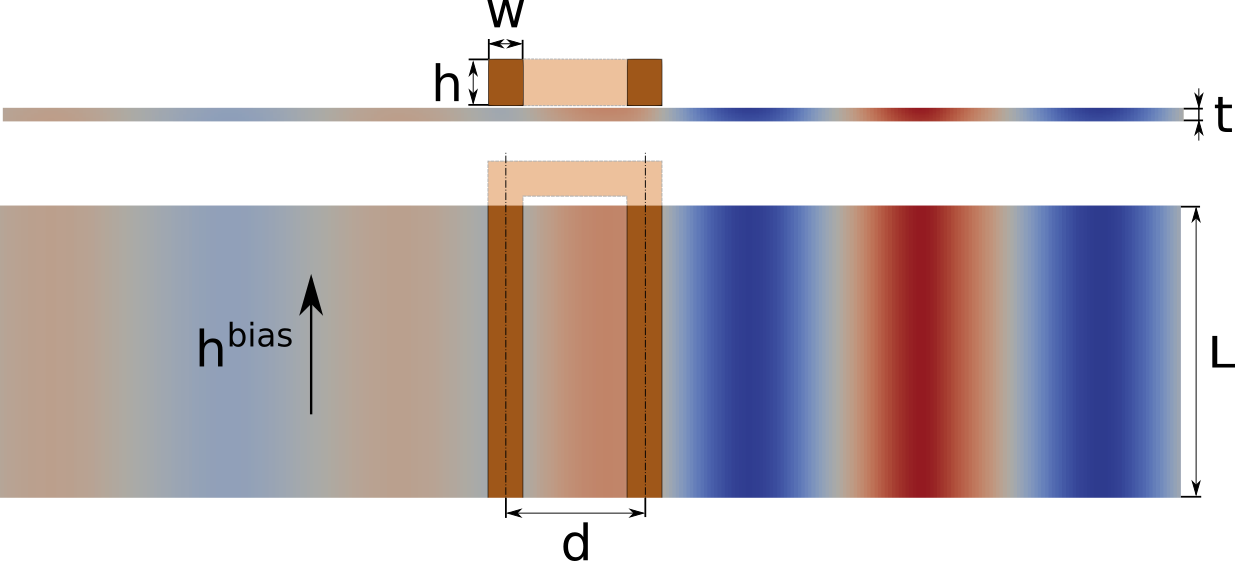}
  \caption{Setup of one U-shaped transducer (brown) consisting of two conductors of length $L$ with a center-to-center distance $d$ and a cross-section $h \times w$. The transducer is placed over a YIG waveguide of thickness $t$ with a small spacing of $\SI{10}{nm}$. For this study the transducer is used in the Damon-Eshbach configuration with an in-plane bias field $\vec{h}^\text{bias}$ orthogonal to the excitation direction. For the sake of simplicity, the simulations neglect parts of the transducer that extend beyond the waveguide. Colors indicate the out-of-plane component of the magnetization $\vec{m}$.  }
  \label{fig:setup}
\end{figure}

A U-shaped geometry shown in Fig.~\ref{fig:setup} is used instead of a coplanar waveguide(CPW) which is placed $\SI{10}{nm}$ above a YIG waveguide in order to supress the spin-pumping effect which might result in the increase of the spin-wave damping \cite{tserkovnyak2002spin}.
As the induced voltage in both branches of the CPW converter might be different, the resulting currents are not divided equally and it would require an additional equation to solve for the correct current distribution. Using an U-shaped transducer avoids this complication and yields insight into optimal design parameters which should also be applicable to the CPW geometry.
The ohmic resistance $R_\Omega$ of the U-shaped transducer is approximated by
\begin{align}\label{eqn:R_omega}
R_\Omega = 2 \, \frac{\rho L}{w h},
\end{align}
with the resistivity $\rho$.
The original transducer dimensions $L_0 \times w_0 \times h_0$ as well as the used material and simulation parameters have been summarized in Table~\ref{tbl:parameters}.

\begin{table}[h!]
  \centering
  \begin{tabular}{|c|c|c|}
  \hline
  \multicolumn{3}{|c|}{\textbf{Transducer (U-shaped)}} \\ \hline
    Dimensions & \( w_0 \times L_0 \times h_0 \) & \( \SI{250}{nm} \times \SI{100}{\mu m} \times \SI{340}{nm} \) \\ \hline
    Distance & \( d \) & \( \SI{1}{\mu m} \) \\ \hline
    Resistivity & \( \rho \) & \( \SI{1.68e-8}{\ohm m}$ (Copper) \\ \hline
  \multicolumn{3}{|c|}{\textbf{Waveguide}} \\ \hline
    Saturation Magnetization & \( \mu_0 M_s \) & \( 138 \ mT \) \\ \hline
    Exchange Constant & \( A \) & \( \SI{3.85}{pJ/m} \) \\ \hline
    Damping Constant & \( \alpha \) & \( 0 \) \\ \hline
  \multicolumn{3}{|c|}{\textbf{Simulation Parameters}} \\ \hline
    Mesh Discretization & \( dx \) & \( \SI{25}{nm} \times \SI{100}{\mu m} \times \SI{10}{nm} \) \\ \hline
    Mesh Elements & \( n \) & \( 2800 \times 1 \times 45 \) \\ \hline
    Operating Frequency & \( f \) & \( \SI{4}{GHz} \) \\ \hline
    Simulation Time & \( T \) & \( \SI{10}{ns} \) \\ \hline
  \end{tabular}
  \caption{Summary of the initial design and material parameters.}
  \label{tbl:parameters}
\end{table}

%
%

In the following subsection several parameter sweeps are performed in order to determine its influence on the transducer efficiency $\eta_T$.
We are focusing on the spin-wave efficiency $\eta_\text{sw}$ and assume that the reflection losses can be nearly perfectly compensated by scaling the transducer length.
Since spin-wave resistance as well as ohmic resistance scale linearly with the length $L$, the length can be chosen to end up with a total resistance $\Re{Z_l} = \SI{50}{\ohm}$. Note, that the inductive reactance $i \omega L_0$ can be neglected for the transducer optimization, since it only leads to a phase-shift and does not directly cause losses.

A single simulation starts with a relaxation of the magnetization inside of the waveguide, followed by the excitation of a spin wave via the Oersted field of a given current distribution.
Simulating $\SI{10}{ns}$ approximately lasts a few minutes on a modern Nvidia A100 GPU and allows to determine the spin-wave resistance for a given bias field.
The bias field influences the spin-wave dispersion inside of the waveguide and has to be chosen in a way that the wavelength $\lambda^\text{sw}$ of the spin wave matches the geometry of the transducer $\lambda^\text{sw} \approx 2 d$.
Thus, for each geometry simulations with different bias fields need to be performed, and only the maximal value of the spin-wave resistance is used.

\subsection{Transducer Height Sweep}
\begin{figure}[h!]
    \centering
    \includegraphics[width=12cm]{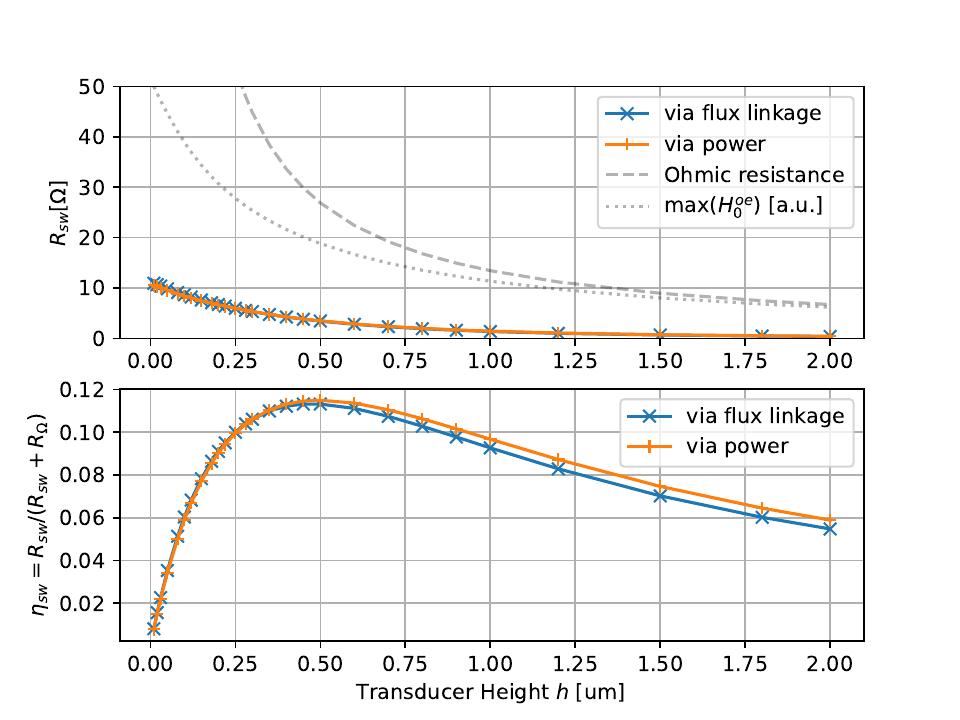}
    \caption{Simulation of the spin-wave resistance $R_\text{sw}$ and spin-wave efficiency $\eta_\text{sw}$ for varying tranducer height $h$. The dashed line indicates the ohmic resistance Eqn.~\eqref{eqn:R_omega}. The dotted line indicates the relative decay of the maximum Oersted field inside of the YIG waveguide.}
    \label{fig:transducer_thickness_sweep}
  \end{figure}

  Increasing the transducer height $h$ reduces the ohmic resistance and thus increases the spin-wave efficiency.
  However, it also increases the average distance from the YIG film and thus reduces the excitation field leading to a drop of the efficiency if the transducer height gets too large.
  The simulated spin-wave resistance and the maximum excitation field, $\max \{\vec{h}^\text{oe}_0\}$, within the YIG waveguide are presented in Fig.~\ref{fig:transducer_thickness_sweep}. The maximum spin-wave efficiency occurs at a transducer thickness of $h \approx \SI{500}{nm}$, but it remains relatively low due to the still substantial ohmic resistance. An additional effect, which limits the potential of increasing the transducer height, is the skin effect, which prevents a uniform current density across the entire cross-section. For example the skin-depth in copper at $f=\SI{4}{GHz}$ is $\delta = \sqrt{\frac{\rho}{\pi \mu f}} \approx \SI{1}{\mu m}$.

\subsection{Transducer Width Sweep}
\begin{figure}[h!]
    \centering
    \includegraphics[width=12cm]{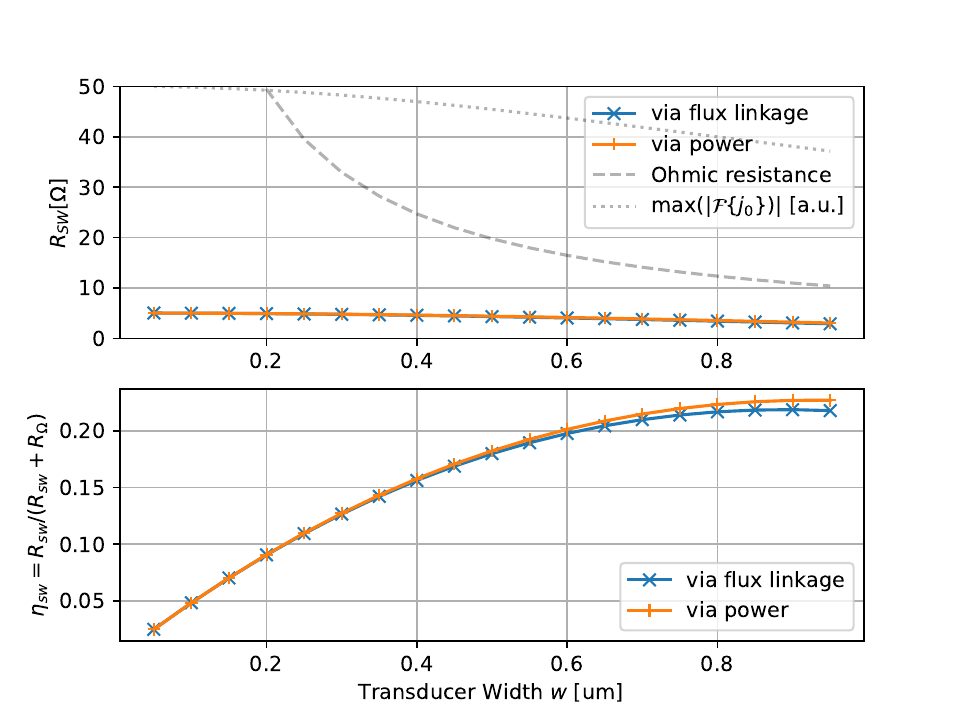}
    \caption{Simulation of the spin-wave resistance $R_\text{sw}$ and spin-wave efficiency $\eta_\text{sw}$ for varying tranducer width $w$. The dashed line indicates the ohmic resistance Eqn.~\eqref{eqn:R_omega}. The dotted line shows the maximum amplitude of the spatial Fourier transform. }
    \label{fig:transducer_width_sweep}
  \end{figure}

  Increasing the transducer width $w$, while maintaining a constant center-to-center distance $d$, provides a straightforward method to reduce ohmic resistance, while the spin-wave resistance remains almost constant. Since the dimensions of the YIG waveguide and also the selected wavelength remain unchanged, the optimal bias field only needs to be determined once. The spin-wave resistance $R_\text{sw}$ and the resulting spin-wave efficiency $\eta_\text{sw}$ for various transducer widths are depicted in Fig.~\ref{fig:transducer_width_sweep}.
  The increase in transducer width has minimal impact on spin-wave resistance. Although, for a given total current $\hat{I}_1$ the current density drops, its spatial Fourier transform $\vert \mathcal{F} \{\vec{j}_0\} \vert$ stays almost constant. Simultaneously, ohmic resistance decreases due to the larger cross-section. Therefore, increasing antenna width is a beneficial strategy, making antennas with wide conductors and narrow gaps between them advantageous. If the gap becomes too small, fabrication challenges may arise, and the increased capacitance between the parallel lines will begin to significantly impact the transducer's performance.

\subsection{Wavelength Sweep}
\begin{figure}[h!]
    \centering
    \includegraphics[width=12cm]{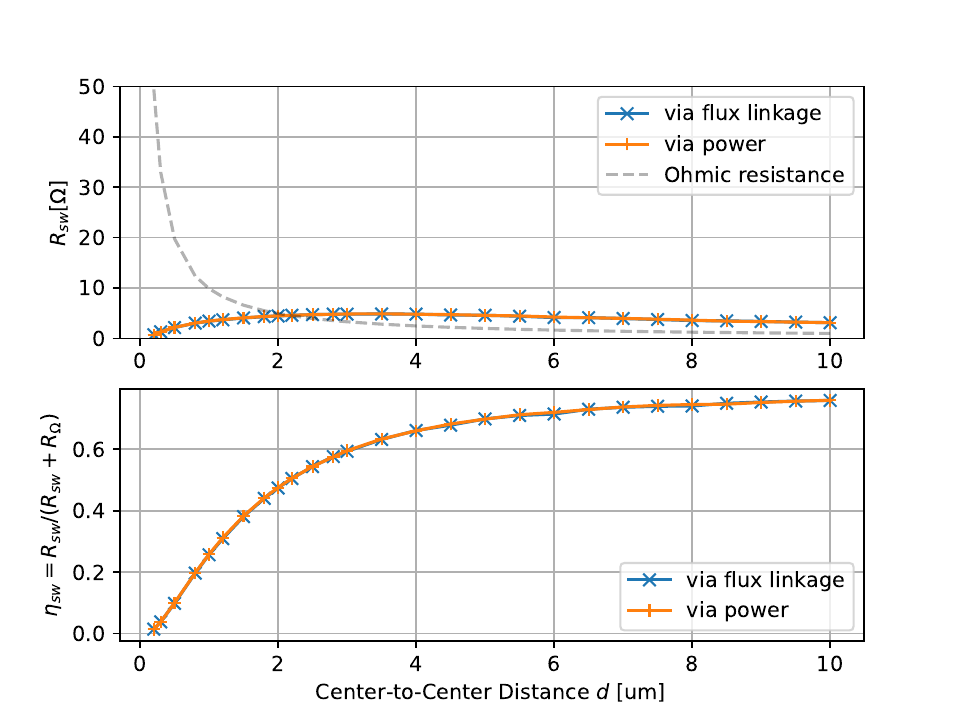}
    \caption{Simulation of the spin-wave resistance $R_\text{sw}$ and spin-wave efficiency $\eta_\text{sw}$ for varying Center-to-Center distance $d$ and constant gap $g = \SI{100}{nm}$ between the two transducer lines. The optimal bias $h^\text{bias}$ has been determined for each distance $d$. The dashed line indicates the ohmic resistance Eqn.~\eqref{eqn:R_omega}. }
    \label{fig:wavelength_sweep}
  \end{figure}

  Increasing the maximal transducer width $w$ is constrained by the center-to-center distance $d$. Increasing the center-to-center distance $d$, which is related to the excited spin-wave wavelength $\lambda^\text{sw} \approx 2 \, d$, allows a further reduction of the ohmic resistance. Simulation results for varying distance $d$ and constant gap $g = d-w = \SI{100}{nm}$ have been summarized in Fig.~\ref{fig:wavelength_sweep}.
  Note, that such optimizations require more computationally intensive simulations, as the optimal bias field must be recalculated for each geometry due to changes in the optimal spin wave wavelength.
  As before, the main effect is the reduction of the ohmic resistance, while the spin-wave resistance remains in the same order of magnitude.

\subsection{Waveguide Thickness Sweep}
\begin{figure}[h!]
    \centering
    \includegraphics[width=12cm]{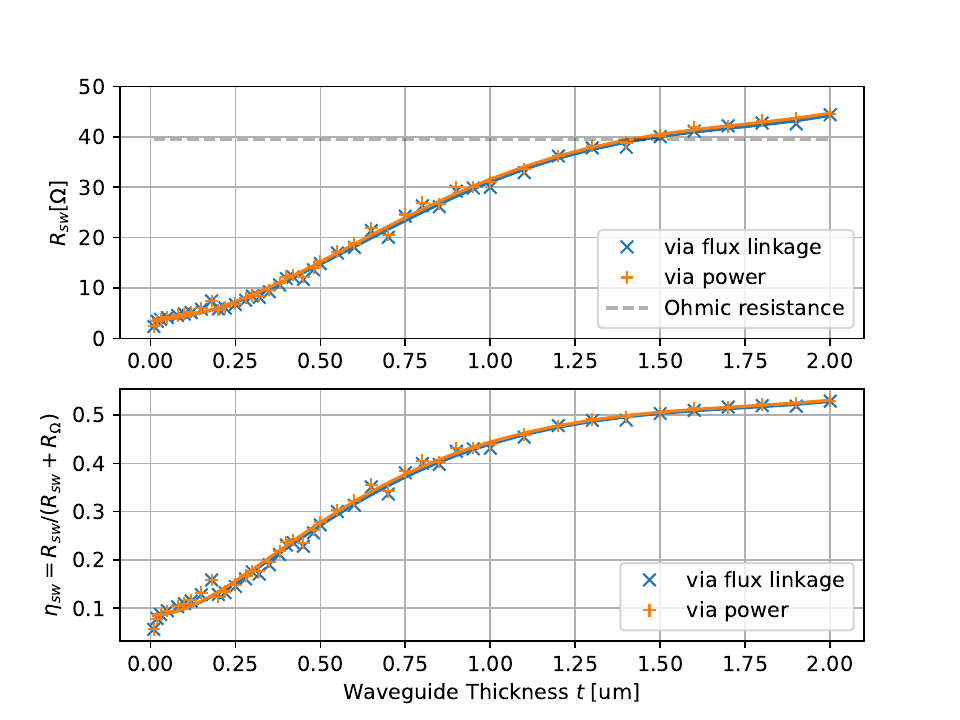}
    \caption{Simulation of the spin-wave resistance $R_\text{sw}$ and spin-wave efficiency $\eta_\text{sw}$ for varying waveguide thickness $t$. The optimal bias $h^\text{bias}$ has been determined for each thickness $t$. The dashed line indicates the ohmic resistance Eqn.~\eqref{eqn:R_omega}. }
    \label{fig:waveguide_thickness_sweep}
  \end{figure}

  While adjusting the transducer dimensions focuses on reducing ohmic resistance, optimizing the YIG waveguide geometry directly effects the spin-wave resistance.
  The results, summarized in Fig.~\ref{fig:waveguide_thickness_sweep}, demonstrate that increasing the waveguide thickness $t$ can substantially enhance the spin-wave resistance.
  Since the ohmic resistance remains constant, this directly improves the spin-wave efficiency.
  Since the spin wave amplitude of magnetostatic surface spin waves shows a $e^{-kz}$ depth dependence \cite{damon1960magnetostatic}, with the wavevector $k = \frac{2\pi}{\lambda^\text{sw}}$, the spin-wave resistance stagnates for waveguides much thicker than the chosen wavelength $\lambda^\text{sw}$.

\section{Conclusion}
Classical spin wave RF devices with insertion loss below \SI{3}{dB} were realized using YIG thicknesses of approximately \SI{30}{\mu m}. \cite{levchenko2024review} However, this technology does not support the miniaturization of RF devices needed for mobile technologies to enable RF filtering for the 5G high band of \SI{26}{GHz} as a replacement for SAW-based technology.
The proposed method for the efficient calculation of the transducer excitation efficiency offers a valuable tool for optimizing the design of spin wave transducers, more applicable to nanoscale devices, such as those reported in \cite{davidkova2025nanoscaled}. By utilizing {\it magnum.np}, all necessary calculations can be seamlessly performed within a single framework. This simplifies the simulation process, enhances accessibility, and enables the adoption of advanced optimization techniques, such as inverse design.

The numerical experiments presented in Section 3 provide critical insights into optimizing spin-wave transducers for higher efficiency. The studies highlight that increasing transducer dimensions, such as height and width, can significantly impact the ohmic resistance and spin-wave excitation efficiency, though practical considerations like fabrication constraints and material properties impose limits. Parameter sweeps on waveguide thickness and wavelength reveal their substantial roles in enhancing spin-wave resistance and overall efficiency.
Although, within this study we focused on the calculation of the spin-wave resistance $R_\text{sw}$ as a key parameter, the method can easily be extended to derive all parameters $Z_{11}$, $Z_{12}$, $Z_{21}$, and $Z_{22}$ of the lumped-circuit model.

\appendix

\section{Acknowledgments}
This research was funded in whole or in part by the Austrian Science Fund (FWF) [10.55776/PAT3864023]. For open access purposes, the author has applied a CC BY public copyright license to any author accepted manuscript version arising from this submission.

\bibliographystyle{ieeetr}
\bibliography{refs}

\end{document}